\documentclass[letter,twocolumn, superscriptaddress,nofootinbib,showkeys,showpacs,10pt]{revtex4}
\usepackage{amssymb}
\usepackage{amsmath}
\usepackage{makeidx}
\usepackage{amsfonts}
\usepackage{graphicx,epstopdf}
\usepackage[ansinew]{inputenc}
\usepackage[usenames,dvipsnames]{pstricks}
\usepackage{subfigure}
\usepackage{pdfpages}
\usepackage{epsfig}
\usepackage{pst-grad}
\usepackage{pst-plot}
\usepackage[colorlinks,hyperindex]{hyperref}

\hypersetup{
colorlinks,
citecolor=blue,
linkcolor=black,
urlcolor=black,
}

\newcommand{\be}{\begin{equation}}
\newcommand{\ee}{\end{equation}}

\DeclareMathOperator{\sech}{sech}
\newcommand{\beq}{\begin{eqnarray}}
\newcommand{\eeq}{\end{eqnarray}}

\begin{document}

\title{Information content in $F(R)$ brane models with non-constant curvature%
}
\author{R. A. C. Correa}
\email{rafael.couceiro@ufabc.edu.br}
\affiliation{CCNH, Universidade Federal do ABC, 09210-580, Santo André, SP, Brazil}
\author{P. H. R. S. Moraes}
\email{moraes.phrs@gmail.com}
\affiliation{ITA - Instituto Tecnol\'ogico de Aeron\'autica - Departamento de F\'isica,
12228-900, S\~ao Jos\'e dos Campos, S\~ao Paulo, Brazil}
\author{A. de Souza Dutra}
\email{dutra@feg.unesp.br}
\affiliation{UNESP, Universidade Estadual Paulista, 12516-410, Guaratinguetá, SP, Brazil}
\author{Roldão da Rocha}
\email{roldao.rocha@ufabc.edu.br}
\affiliation{CMCC, Universidade Federal do ABC, 09210-580, Santo André, SP, Brazil}
%\author{R. Menezes}
%\email{menezes@ufpb.br}
%\affiliation{Departamento de Ciências Exatas, Universidade Federal da Paraíba, 58297-000,
%Rio Tinto, PB, Brazil}
%\affiliation{Departamento de Física, Universidade Federal de Campina Grande, 58109-970,
%Campina Grande, PB, Brazil}

\begin{abstract}
In this work we investigate the entropic information-measure in the context
of braneworlds with non-constant curvature. The braneworld entropic
information is studied for gravity modified by the squared of the Ricci
scalar, besides the usual Einstein-Hilbert term. We showed that the minimum
value of the brane configurational entropy provides a stricter bound on the
parameter that is responsible for the $F(R)$ model to differ from the
Einstein-Hilbert standard one. Our results are moreover consistent to a
negative bulk cosmological constant.
\end{abstract}

\maketitle

\section{Introduction}

The standard model of cosmology, named $\Lambda$CDM model, derived from
Einstein's General Relativity, although yields a great accordance between
theory and observational data \cite{hinshaw/2013}, has some shortcomings
which question its validity as the truthful model for the origin, structure
and evolution of the Universe. Among those shortcomings, one could quote the
cosmological constant (CC), the coincidence and the dark matter problems,
missing satellites, hierarchy problem and others likewise (see \cite%
{clifton/2012} and references therein).

The CC problem is the most critical among those issues, since it consists of
a lack of convincing explanation for the physical meaning of dark energy,
which composes $\sim70\%$ of the Universe and, in principle, is responsible
for the cosmic acceleration predicted by Type Ia Supernovae observational
data \cite{riess/1998,perlmutter/1999}.

In order to evade some of those shortcomings, it is common to consider
generalized theories of gravity, such as the $F(R)$ theories (check \cite%
{defelice/2010,sotiriou/2010} for instance), as the starting point for
alternative cosmological models. Such a formalism successfully describes
both the inflationary era \cite{cognola/2008,huang/2014} and the current
phase of accelerated expansion our Universe is undergoing \cite%
{oikonomou/2013,starobinsky/2007}, the latter, with no need of a CC.

On the other hand, the hierarchy problem, for instance, may be solved from
the approach of braneworld models \cite{randall/1999,sahni/2003}. This
occurs since in such a Universe set up, gravity is allowed to propagate
through the bulk (a five dimensional anti-de Sitter space-time), differently
from the other fundamental forces of nature. This explains the ``weakness"
of gravity in the observable Universe.

Note that important outcomes also raise from the approach of generalized $%
F(R)$ gravity in braneworld models. In \cite%
{Bazeia-Lobao-Menezes-Petrov-Silva}, for instance, the authors obtained
exact solutions for the scalar field, warp factor and energy density in a
scenario with non-constant curvature. Analytical solutions for the equations
of motion in the case of constant curvature were presented in \cite%
{Afonso-Bazeia-Menezes-Petrov}. The modified Einstein's equations were
solved for a flat brane in \cite{bazeia/2013}. Furthermore, cosmological
solutions for a fourth-order $F(R)$ brane gravity are presented in \cite%
{balcerzak/2010}. For other works on $F(R)$ branes, check \cite%
{parry/2005,deruelle/2008,hoffdasilva/2011,xu/2015,liu/2011}.

Despite the amount of applications at which $F(R)$ brane models have been
applied recently, no efforts have been accomplished yet in the framework of
the so-called configurational entropy (CE) in these scenarios. Gleiser and
Stamatopoulos (GS) have firstly proposed in \cite{Gleiser-Stamatopoulos}
such a new physical quantity, which brings additional informations about
some parameters of a given model for which the energy density is localized.
It has been shown that the higher the energy that approximates the actual
solution, the higher its relative CE, which is defined as the absolute
difference between the actual function CE and the trial function CE. As
pointed out in \cite{Gleiser-Stamatopoulos}, the CE can resolve situations
where the energies of the configurations are degenerate. In this case, the
CE can be used to select the best configuration.

The approach presented in \cite{Gleiser-Stamatopoulos} has been used to
study the non-equilibrium dynamics of spontaneous symmetry breaking \cite%
{PRDgleiser-stamatopoulos}, to obtain the stability bound for compact
objects \cite{Gleiser-Sowinski}, to investigate the emergence of localized
objects during inflationary preheating \cite{PRDgleiser-graham}, and
moreover, to distinguish configurations with energy-degenerate spatial
profiles \cite{Rafael-Dutra-Gleiser}. Furthermore, in a recent work \cite%
{Rafael-Roldao}, solitons, Lorentz symmetry breaking, supersymmetry, and
entropy, were employed using the CE concept. In such a work, the CE for
travelling solitons reveals that the best value of the parameter responsible
for breaking the Lorentz symmetry is $1$ where the energy density is
distributed equally around the origin. In this way, it was argued that the
information-theoretical measure of travelling solitons in Lorentz symmetry
violation scenarios can be very important to probe situations where the
parameters responsible for breaking the symmetries are arbitrary. In this
case, the CE was shown to select the best value of the parameter in the
model. Another interesting work about CE was presented in Ref. \cite%
{stamatopoulos/2015}, where the CE is responsible for identifying the
critical point in the context of continuous phase transitions. Finally, in
braneworld scenarios \cite{bc} it was shown that CE can be employed to
demonstrate a high organizational degree in the structure of the system
configuration for large values of a parameter of the sine-Gordon model.

\textcolor{black}{In this work we are interested in answering the following issues. Can
the CE be calculated in $F(R)$ brane scenarios? If it does, how is its profile? Furthermore, what the information content
in $F(R)$ brane models with non-constant curvature may reveal?}

\textcolor{black}{We will show that the
CE provides a stricter bound on the parameter that is responsible for the $F(R)$ model to differ from the standard gravity one.}

\textcolor{black}{This paper is organized as follows. In the next section, we present a brief
review of $F(R)$ brane models. In particular, we review the results presented by
Bazeia and collaborators \cite{Bazeia-Lobao-Menezes-Petrov-Silva}. In Sec.III, we present an overview regarding CE measure, and
we calculate the entropic information for the $F(R)$ brane models. In Sec.IV, we
show a comparison between the results of the
information-entropic measure of $F(R)$ brane models and what is obtained via cosmology.  In Sec.V, we present our
conclusions and final remarks.}

\section{A brief review of $F(R)$ brane models}

In this section a brief overview regarding $F(R)$ braneworld models is
presented. Let us start by writing the action of five-dimensional gravity
coupled to a real scalar field $\phi $ as 
\begin{equation}
S\!=\!\int d^{5}x\sqrt{|g|\,}\left[ -\frac{1}{4}F(R)+\frac{1}{2}g^{ab}\nabla
_{a}\phi \nabla _{b}\phi -V(\phi )\right].  \label{eq.1}
\end{equation}
Here $4\pi G_{(5)}\!=\!1$ and $g\!=\!det(g_{ab})$, with field, space and
time variables being dimensionless. $F(R)$ stands for a generic function of
the Ricci scalar $R$. Furthermore, the signature of the metric is adopted as 
$(+----)$. It should be stressed that $V(\phi )$ is the potential that
describes the theory.

We will study the case where the metric is represented by 
\begin{equation}
ds^{2}\!=\!e^{2A}\eta _{\mu \nu }dx^{\mu }dx^{\nu }-dy^{2},  \label{eq3}
\end{equation}%
where $y$ denotes the extra dimension, $\eta _{\mu \nu}$ is the usual
Minkowski metric, and $e^{2A}$ stands for the so-called warp factor, which
depends only on the extra dimension. Moreover, let us assume that the field $%
\phi$ also depends solely upon $y$. Hence, from the action (\ref{eq.1}) the
corresponding equation of motion for the scalar field reads 
\begin{equation}
\phi ^{\prime \prime }+4A^{\prime }\phi ^{\prime }\!=\!V_{\phi },
\label{eq.4}
\end{equation}%
wherein the primes stand for derivatives with respect to the extra dimension
and $V_{\phi }\!=\!dV/d\phi $.

Here, the energy density $\rho $ is given by 
\begin{equation}
\rho \!=\!-e^{2A}\mathcal{L},  \label{eq.5}
\end{equation}
\noindent with%
\begin{equation}
\mathcal{L}=\frac{1}{2}g^{ab}\nabla _{a}\phi \nabla _{b}\phi -V(\phi ).
\label{eq.6}
\end{equation}

Now, after straightforward manipulations the modified Einstein's equations
acquire the form 
\begin{eqnarray}
-\frac{2}{3}\phi ^{\prime 2} &=&-\frac{1}{3}A^{\prime \prime }F_{R}^{\prime
}+\frac{1}{3}F_{R}^{\prime \prime }+A^{\prime \prime }F_{R},  \label{eq.7} \\
{} -\frac{1}{2}\phi ^{\prime 2}\!+\!V(\phi ) &=&2(A^{\prime \prime
}\!+\!A^{\prime 2})F_{R}\!-\!2A^{\prime }F_{R}^{\prime }\!-\!\frac{1}{4}F,
\label{eq.8}
\end{eqnarray}
\noindent with $F_{R}:=dF/dR$. The Ricci scalar is assumed to be an
arbitrary function of the extra dimension, i.e., $R\!=\!R(y)$, yielding from
Eq.(\ref{eq.7}) that 
\begin{equation}
\phi ^{\prime 2}\!=\!-\frac{3}{2}A^{\prime \prime }F_{R}\!+\!\frac{1}{2}%
\left( A^{\prime }R^{\prime }\!-\!R^{\prime \prime }\right) F_{RR}\!-\!\frac{%
1}{2}F_{RRR}R^{\prime 2}.  \label{eq.91}
\end{equation}
It is worth to mention that $\lim_{y\to0}R(y) = 20B^2k^2$, where $%
\lim_{y\to\infty}R(y)=8Bk^2$ \cite{Afonso-Bazeia-Menezes-Petrov}. The
potential can be obtained also from (\ref{eq.8}): 
\begin{eqnarray}
&&\left. V(\phi )=-\frac{1}{4}F+\frac{1}{4}F_{R}(8A^{\prime 2}+5A^{\prime
\prime })\right.  \notag \\
&&\left. -\frac{1}{4}F_{RR}(A^{\prime \prime }+7A^{\prime }R^{\prime })-%
\frac{1}{4} F_{RRR}R^{\prime 2}.\right.  \label{eq.10}
\end{eqnarray}
Hence, by substituting the equation $R=8A^{\prime \prime }+20A^{\prime 2}$
for the Ricci scalar into Eqs.(\ref{eq.91}) and (\ref{eq.10}), yields 
\begin{eqnarray}
\phi ^{\prime 2}\!&=&\!-\frac{3}{2}F_{R}A^{\prime \prime }\!+4F_{RR}\left(
5A^{\prime 2}A^{\prime \prime }-5A^{\prime \prime 2}-4A^{\prime }A^{\prime
\prime \prime }\right.  \notag \\
&&\left. \left. -A^{\prime \prime \prime \prime }\right)
-32F_{RRR}(5A^{\prime }A^{\prime \prime }+A^{\prime \prime \prime
})^{2},\right.  \label{eq.12} \\
{} V(\phi )&=&-\frac{1}{4}F+\frac{1}{4}F_{R}(5A^{\prime \prime }+8A^{\prime
2})  \notag \\
{} &&\left. -F_{RR}(70A^{\prime 2}A^{\prime \prime }+10A^{\prime \prime
2}+24A^{\prime \prime \prime }A^{\prime }+A^{\prime \prime \prime \prime
})\right.  \notag \\
{} &&\left. -16F_{RRR}(5A^{\prime }A^{\prime \prime }+A^{\prime \prime
\prime })^{2}.\right.  \label{eq.13}
\end{eqnarray}
In order to explicit find solutions for the above equations, the function $%
F(R)\!=\!R+\alpha R^{2}$ can be employed \cite%
{Afonso-Bazeia-Menezes-Petrov,liu/2011}, where $\alpha\in\mathbb{R}$. Hence
Eqs.(\ref{eq.12}) and (\ref{eq.13}) can be recast as 
\begin{eqnarray}
\phi ^{\prime 2}\!&=&-\frac{3}{2}A^{\prime \prime }-4\alpha \left(
16A^{\prime \prime 2}+5A^{\prime 2}A^{\prime \prime } +2A^{\prime \prime
\prime \prime }+8A^{\prime }A^{\prime \prime \prime }\right) ,  \notag \\
{} V(\phi )&=&-\frac{3}{4}A^{\prime \prime }-3A^{\prime 2}-2\alpha\left(
10A^{\prime 4}+24A^{\prime }A^{\prime \prime \prime }\right.  \notag \\
{} && \left. +8A^{\prime \prime 2}+2A^{\prime \prime \prime \prime
}+69A^{\prime 2}A^{\prime \prime }\right) .  \label{eq.15}
\end{eqnarray}%
Moreover, Eq.(\ref{eq.5}) can be expressed as 
\begin{eqnarray}
\rho(y) &=&e^{2A}\left[ \frac{3}{2}A^{\prime \prime }+3A^{\prime 2}+4\alpha
(5A^{\prime 4}+16A^{\prime }A^{\prime \prime \prime }\right.  \notag \\
{} &&\left. 37A^{\prime 2}A^{\prime \prime }+12A^{\prime \prime
2}+2A^{\prime \prime \prime \prime })\right] ,  \label{eq.16}
\end{eqnarray}
what implies that 
\begin{equation}
E=\frac{44\alpha }{3}\int dy\,A^{\prime 4}\,e^{2A}\,.  \label{eq.16.1}
\end{equation}
Now, in order to work with analytical solutions, the warp function is
adopted to be \cite{Gremm} 
\begin{equation}
A(y)\!=\!B \ln \left[ \sech(ky)\right] ,  \label{eq11}
\end{equation}%
where $B>0$ and $k>0$. Hence the energy density reads \cite%
{Bazeia-Lobao-Menezes-Petrov-Silva} 
\begin{equation}
\rho (y)\!=\!g_{1}\mathrm{sh}^{2B}(ky)\!+\!g_{2}\mathrm{sh}%
^{2B+2}(ky)\!+\!g_{3}\mathrm{sh}^{2B+4}(ky),  \label{3.4}
\end{equation}%
with sh $\equiv \sech$ and 
\begin{eqnarray}
g_{1} &&:=-3B^{2}k^{2}(20\alpha B^{2}k^{2}+1),  \notag \\
{} g_{2} &&:=3Bk^{2}\!\left[4\alpha
k^{2}\left(10B^{3}\!+\!37B^{2}\!+\!32B\!+\!8\right) +B\!+\!\frac{1}{2}\right]
,  \notag \\
{} g_{3} &&:=-12\alpha k^{4}B\left( 5B^{3}+37B^{2}+44B+12\right) .
\end{eqnarray}
The parameter $\alpha $ is bounded by 
\begin{equation}
{32k^{2}(1\!+\!4B)}\leq\!3\alpha^{-1} \!\leq-{8k^{2}(8\!+\!16B\!+\!5B^{2})}%
\,.  \label{eq15}
\end{equation}

Now, substituting the warp function (\ref{eq11}) into (\ref{eq.16.1}), we
have the energy $E$ of the brane: 
\begin{equation}
E=11\sqrt{\pi }\alpha k^{3}{B^{4}\Gamma (B)}/{\Gamma (B+5/2)}\,.
\label{eq.18}
\end{equation}

\textcolor{black}{Thus, in the next section, we will use the approach presented here to obtain
the CE in this context. As we will see, the
information-entropic measure shows a higher organizational degree in the
structure of the system configuration, and consequently we will be able to obtain additional
information content regarding the system regarded.}

\section{Information Content in $F(R)$ Brane Models}

As argued in the Introduction, GS have recently proposed a detailed picture
of the so-called Configurational Entropy for the structure of localized
solutions in classical field theories \cite{Gleiser-Stamatopoulos}.
Analogously to that development, we present a CE measure in functional
space, from the field configurations where braneworld models can be studied.
The framework is going to be here revisited and subsequently applied.

{{There is an intimate link between information and dynamics, where the
entropic measure plays a prominent role. The entropic measure is well known
to quantify the informational content of physical solutions to the equations
of motion and their approximations, namely, the CE in functional space \cite%
{Gleiser-Stamatopoulos}. GS proposed that nature optimizes not solely by
extremizing energy through the plethora of \emph{a priori} available paths,
but also from an informational perspective. }}

To start, let us write the following Fourier transform%
\begin{equation}
\mathcal{F}[\omega ]=\frac{1}{\sqrt{2\pi }}\int dy\rho (y)e^{i\omega y}.
\label{3.1}
\end{equation}
%From the Plancherel theorem it follows that

%\begin{equation}
%\int d\omega \left\vert \mathcal{F}[\omega ]\right\vert ^{2}=\int
%dy\left\vert \rho (y)\right\vert ^{2}.  \label{3.1.1}
%\end{equation}

Now, the modal fraction measures the relative weight of each mode $\omega $
and is defined by expression \cite{Gleiser-Stamatopoulos, Gleiser-Sowinski,
Rafael-Dutra-Gleiser, Rafael-Roldao}: 
\begin{equation}
f(\omega )=\frac{\left\vert \mathcal{F}[\omega ]\right\vert ^{2}}{\int
d\omega \left\vert \mathcal{F}[\omega ]\right\vert ^{2}}.  \label{3.2}
\end{equation}

{The CE is inspired by the Shannon's information framework, being defined by 
}$\mathfrak{S}${{$_{C}[f]=-\sum f_{n}\,\ln (f_{n}) $. It represents an
absolute limit on the best lossless compression of communication \cite%
{Shannon}. Hence the CE at first provided the informational lining regarding
configurations which are compatible to constraints of arbitrary physical
systems. When $N$ modes labeled by $k$ carry the same weight, it follows
that $f_{n}=1/N$ and the discrete CE presents a maximum value at }}$%
\mathfrak{S}${{$_{C}=\ln N$, accordingly. Alternatively, if the system is
embodied by merely one mode, consequently }}$\mathfrak{S}${{$_{C}=0$ \cite%
{Gleiser-Stamatopoulos}.}}

{Similarly, for arbitrary non-periodic functions in an open interval, the
continuous CE reads} 
\begin{equation}
\mathfrak{S}_{c}[f]=-\int d\omega \mathring{f}(\omega )\ln [\mathring{f}%
(\omega )],  \label{3.3}
\end{equation}%
\noindent where $\mathring{f}(\omega ):=$ $f(\omega )/f_{\max }(\omega )$ is
defined as the normalized modal fraction, whereas the term $f_{\max }(\omega
)$ stands for the maximum fraction. Hence, Eq.(\ref{3.1}) engenders the
modal fraction to achieve the entropic profile of thick brane solutions. It
is worth to remark that Eq.(\ref{3.1}) differs from that provided by GS. In
this framework we include the warp factor in $\mathcal{F}[\omega ]$. Hence
the framework brings further information concerning a warped geometric
scenario.

Here, as a straightforward example, we shall calculate the entropic
information for the $F(R)$ model. First the modal fraction can be computed.
First, Eq.(\ref{3.1}) reads 
\begin{eqnarray}
\mathcal{F}[\omega ]=\sum_{m=1}^{2}\!\sum_{j=1}^{3}A_{j,m}\!\times _{2}%
\mathcal{G}_{1}^{(j,m)}\!\left[ \gamma _{j},\mu _{j,m};\mu _{j,m}\!+\!1;-1%
\right]  \label{eq.3.7}
\end{eqnarray}%
\noindent where $_{2}\mathcal{G}_{1}[\;\otimes \;,\;\circledcirc
\;;\;\circledast \;;\;\oplus \;]$ stands for the well-known hypergeometric
functions with 
\begin{eqnarray}
\mu _{j,m} &:&=\frac{1}{2k}(\gamma _{j}k-i(-1)^{m+1}\omega)\,.
\end{eqnarray}
Moreover, $A_{j,m}$ and $\gamma _{j}$ are defined as 
\begin{eqnarray}
A_{j,m} &:&=\frac{1}{\sqrt{2\pi }}\frac{2^{\gamma _{j}-1}g_{j}(\gamma
_{j}k+i(-1)^{m+1}\omega )}{(\omega ^{2}+\gamma _{j}^{2}k^{2})}\,, \\
{} \!\!\!\gamma _{j} &:&=2(B+j-1)\,.  \label{3.8}
\end{eqnarray}

Thus, the modal fraction (\ref{3.2}) becomes 
\begin{equation}
f(\omega )\!=\!\frac{\sum_{p,m=1}^{2}\sum_{q,j=1}^{3}A_{j,m}A_{q,p}^{\ast
}(_{2}\mathcal{G}_{1}^{(j,m)})(_{2}\mathcal{G}_{1}^{(p,q)})^{\ast }}{%
\sum_{p,m=1}^{2}\sum_{q,j=1}^{3}\int d\omega A_{j,m}A_{q,p}^{\ast }(_{2}%
\mathcal{G}_{1}^{(j,m)})(_{2}\mathcal{G}_{1}^{(p,q)})^{\ast }}.  \label{3.9}
\end{equation}

In Fig.1 below the modal fraction is depicted for different values of $%
\alpha $. The maximum of the distributions are localized around the mode $%
\omega =0$. By taking into account the modal fraction in (\ref{3.9}) and its
maximum contribution, Eq.(\ref{3.3}) can now be solved in order to obtain
the Brane Configurational Entropy (BCE). In this case, due to the high
complexity of integration, Eq.(\ref{3.3}) must be integrated numerically.
The results are shown in Fig. 2, where the BCE is plotted as a function of $%
\alpha $. By using a recent approach presented by GS \cite%
{Gleiser-Stamatopoulos}, the BCE is correlated to the energy of the system,
in the sense that the lower [higher] the BCE, the lower [higher] the energy
of the solutions. Moreover, BCE further provides independent criteria to
control the stability of configurations based upon the informational content
of their profiles \cite{Gleiser-Sowinski}. In fact, the BCE maximum
represents the boundary between stability and instability, as the case
analysed in \cite{Gleiser-Sowinski} for $Q$-balls.

In the last section we shall provide the consequences of the model studied
above and point forthcoming perspectives.

\begin{figure}[h]
\begin{center}
\includegraphics[width=8.0cm]{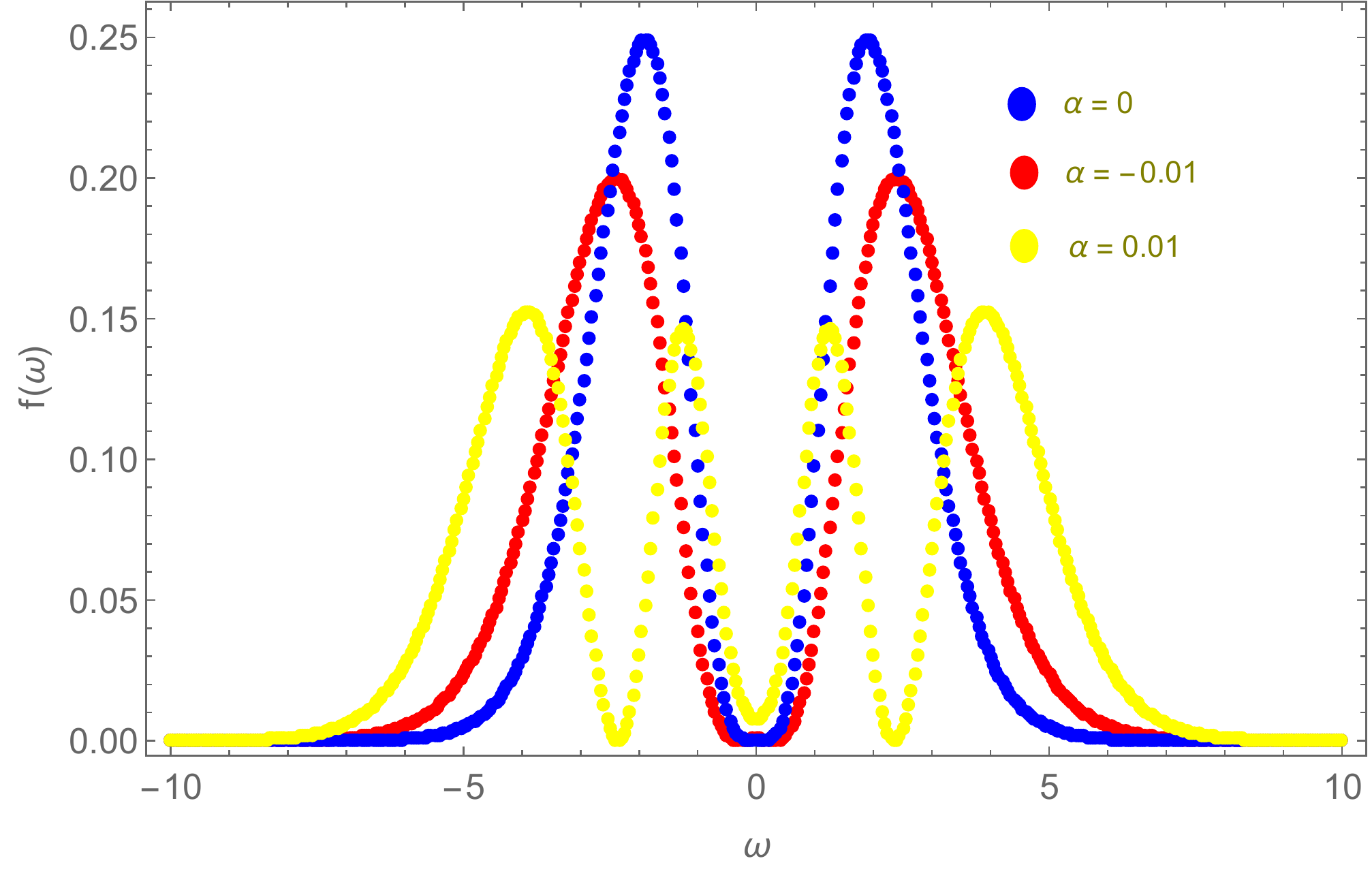}
\end{center}
\caption{Modal fractions for $\protect\alpha = 0$ (blue), $\protect\alpha =
-0.01$ (red) and $\protect\alpha = 0.01$ (yellow). }
\label{fig1:Fractions}
\end{figure}
%%%%%%%%%%%%%%%%%%%%%%%%%%%%%%%%%%%%%555
\begin{figure}[h]
\begin{center}
\includegraphics[width=8.0cm]{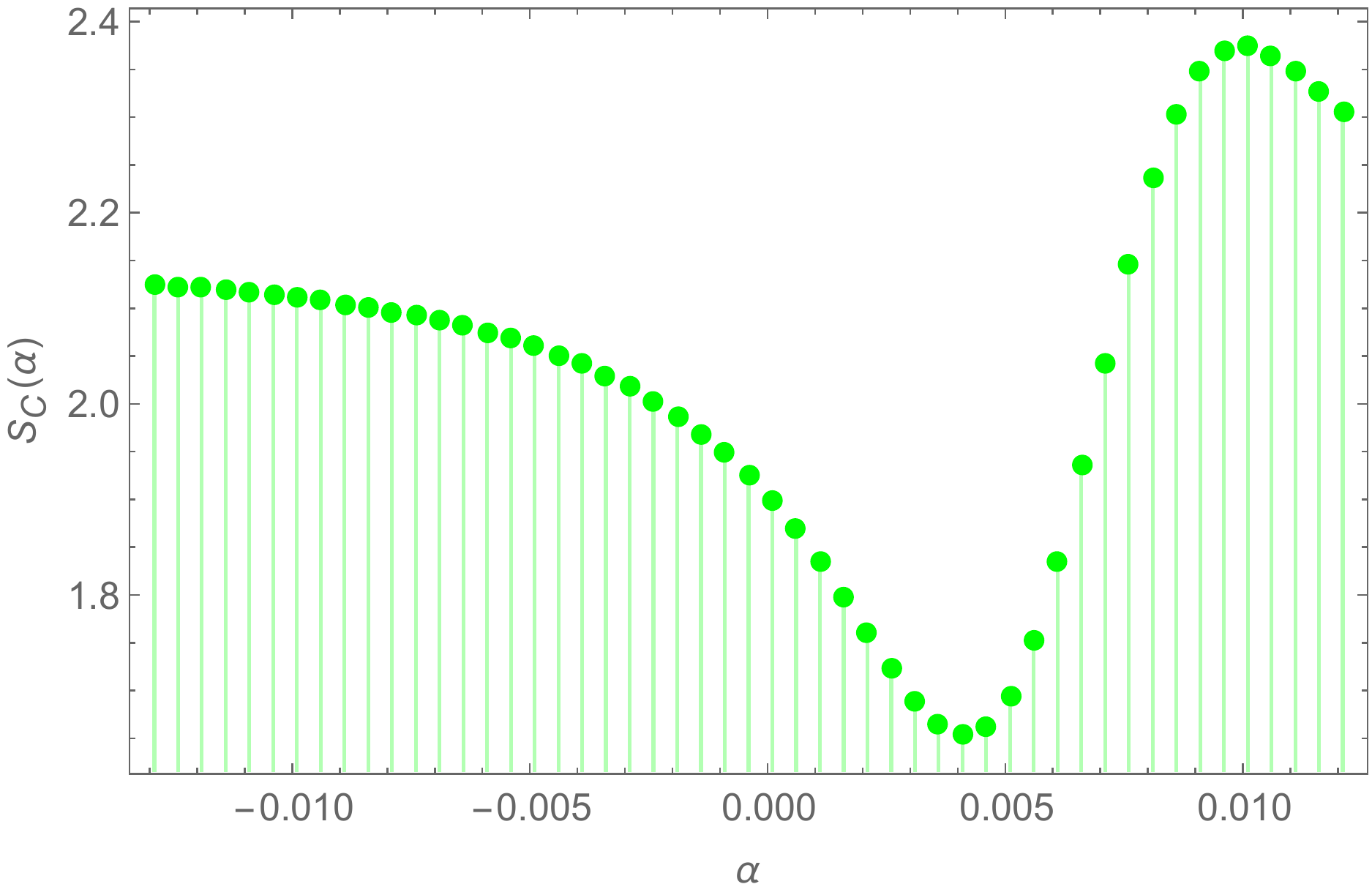}
\end{center}
\caption{Configurational entropy as a function of the parameter $\protect%
\alpha$.}
\label{fig2:Entropy}
\end{figure}

\section{Comparison with cosmology}

\label{cwc}

Cosmological models have been constantly derived from higher order
derivative gravity theories.

For instance, in \cite{liu/2011} the authors have used the same functional
form used in the present paper for $F(R)$, i.e., $F(R)=R+\alpha R^{2}$, to
derive analytical solutions for both the warp factor and scalar field as
functions of $y$. As solutions for some cosmological parameters, as the bulk
CC, the authors have found $\Lambda_5=477/(-6728\alpha\kappa_5^{2})$, with $%
\kappa_5$ representing the five-dimensional coupling constant. Moreover,
another approach which provides similar results predicts that $%
\Lambda_5=-B^2k^2(3 +20\alpha B^2k^2)$ \cite%
{Bazeia-Lobao-Menezes-Petrov-Silva}. Note that in order to obtain a negative
bulk CC, $\alpha$ must be positive, which is in accordance with what was
developed in the previous section. Note also that the negative bulk CC is
the responsible for gravity ``to leak" from the brane to the extra
dimension, but still remain concentrated in our observable universe. On the
other hand, a positive bulk CC would accelerate such a process of ``leaking"
(check \cite{maartens/2010}).

Moreover, in \cite{liu/2011}, for the brane tension, it was found $%
\lambda=3\kappa_5^{2}/(784\alpha)$. Such a relation reinforces the positive
sign of $\alpha$, since a negative tension brane is gravitationally unstable
by itself (check \cite{hoffdasilva/2011}).

Furthermore, an equation which leads to the singularities of the effective
potential on the brane has been constructed (Eq.(48) of \cite{liu/2011}).
Such an equation has solutions only when $\alpha\gtrsim1/(40B^{2})$ with $%
B>2 $. Note that by taking $B=2.5$ in the latter relation, one obtains
exactly the value of $\alpha$ derived via the study of the BCE presented
above, i.e., $\alpha=0.0046$. Indeed, the parameter space of $(B,\alpha)$
was analyzed in \cite{liu/2011}, with the upper limit of $\alpha$ lying on $%
\alpha=0.005$.

In \cite{Bazeia-Lobao-Menezes-Petrov-Silva}, the allowed region of $\alpha$
for distinct values of $B$ was also depicted (check the upper panel of Fig.2
on such a reference). The result covers $\alpha=0.0046$ with the energy
density having a maximum at $y=0$.

\section{Concluding Remarks and Outlook}

The entropic information has been studied in braneworld models, with
emphasis on the $F(R)$ model, which has been chosen by its very physical
content and usefulness. The BCE is moreover exerted to evince a higher
organisational degree in the structure of system configuration likewise. The
GS technique was employed to achieve a correlation involving the energy of
the system and its BCE. Moreover, our analysis is further based upon the CE $%
\mathfrak{S}(\alpha )$, depicted in Fig.2. Such configurations for $\alpha
\simeq 0.0046$ are most probable to be found by the system. In fact, in such
range of $\alpha $ the CE $\mathfrak{S}(\alpha ) $ approaches to zero. Our
results are consistent to the upper limit $\alpha\lesssim0.005$ in \cite%
{liu/2011}, and further imposes the value of $\alpha$ corresponding to the
best ordering from the BCE point of view. Such a value for $\alpha$ was
supported by results obtained purely via $F(R)$ brane cosmological models,
as it can be realized in Section \ref{cwc}.

Once developed the formalism of the BCE and the entropic information as
well, we can further apply a procedure similar to what has been studied in
the previous sections to other thick braneworld models.

\acknowledgments RACC thanks CAPES for financial support. PHRSM thanks
FAPESP for financial support. ASD is thankful to the CNPq for financial
support. RdR thanks CNPq Grants No. 303027/2012-6 and No. 473326/2013-2, and
FAPESP No. 2015/10270-0, for partial financial support.

\bigskip

\end{document}